\documentclass[useAMS,usenatbib]{mn2e}
\usepackage{graphics,rotating,multirow,bm,color,hyperref,amsmath}
\hypersetup{
    colorlinks=true,
    linkcolor=black,
    citecolor=black,
}

\title[Environments in excursion set theory for modified gravity]
  {Excursion set theory for modified gravity: Eulerian versus Lagrangian environments}
\author[Baojiu~Li and Tsz Yan Lam]
  {Baojiu~Li$^{1,2}$\thanks{E-mail: baojiu.li@durham.ac.uk}, Tsz Yan Lam$^2$\thanks{E-mail: tszyan.lam@ipmu.jp}\\
  $^1$Institute of Computational Cosmology, Department of Physics, Durham University, South Road, Durham DH1 3LE, UK\\
  $^2$Kavli-IPMU, University of Tokyo, Kashiwa, Chiba 277-8583, Japan}
  
\pagerange{\pageref{firstpage}--\pageref{lastpage}} \pubyear{2012}

\def\LaTeX{L\kern-.36em\raise.3ex\hbox{a}\kern-.15em
    T\kern-.1667em\lower.7ex\hbox{E}\kern-.125emX}

\newcommand{\tcr}[1]{\textcolor{black}{#1}}

\begin{document}

\label{firstpage}

\maketitle

\begin{abstract}
We have revisited the extended excursion set theory in modified gravity models,
 taking the chameleon model as an example. 
Instead of specifying their Lagrangian size, 
here we define the environments by 
the Eulerian size, 
chosen to be of the same order of the Compton length of the scalar field 
by physical arguments. 
We find that the Eulerian and Lagrangian environments have very different 
environmental density contrast probability distributions, 
the former more likely to have high matter density, 
which in turn suppressing the effect of the fifth force 
 in matter clustering and halo formation. 
The use of Eulerian environments also evades the 
unphysical restriction of having an upper mass limit 
in the case of Lagrangian environments.
Two methods of computing the unconditional mass functions, numerical integration and Monte Carlo simulation, are discussed and 
found to give consistent predictions. 
\end{abstract}


\section{Introduction}

\label{sect:intro}

The excursion set theory \citep{bcek} is a concise yet very successful approach \citep{zentner2007} to study the nonlinear structure formation in the standard cold dark matter (CDM) scenario. Starting from an (usually Gaussian) initial distribution of the density perturbation field at early times, 
with an evolution model such as the spherical collapse model,
 \footnote{In this work we only consider the spherical collapse model and focus on the effect of different definitions of environment on the halo mass function in modified gravity models}
to map an initial density perturbation to a nonlinear structure 
(dark matter halo) at some late time, 
it can predict statistically what fraction of matter has been assembled into the halo at that time. It makes the physics of structure formation clear and simple, and was a major tool for studying the large-scale structure formation when large cosmological simulations were still beyond the capability of supercomputers.

With the coming of the era of precision cosmology and progresses in supercomputing, $N$-body simulations \citep{bertschinger1998} have become more common nowadays for its ability to capture non-linear evolution without any assumptions
in the evolution model and hence 
more accurate predictions than the excursion set theory. Despite this, the analytical results of the latter still provide valuable information. For example, fitting formulae of the mass function can be obtained 
\citep{st2002}, the parameters of which are then calibrated by the numerical simulations \citep{st1999,jetal2001,tetal2008}.

In many aspects, the study of the cause of the accelerated cosmological expansion \citep{cst2006} follows the same history of the enquiries of CDM: people first ask about their implications in the large, linear structures, and then gradually shift towards the smaller, nonlinear scales, in which process ever advanced techniques and tools are developed. 
After an initial burst of theoretical or phenomenological models following the first observational evidences of the cosmic acceleration \citep{riess1998,perlmutter1999}, people have spent years trying to understand the behaviour of all these models on large scales, paving the road that leads to a better and full understanding of the theories and preparing for the confrontations with future data.

One important class of such models involves modifications to Einstein's General Relativity (GR) on large scales \citep{jk2010,cfps2011}. Although these modified gravity models are mainly designed to explain the observations, many of the ideas are motivated from studies in fundamental physics, making them very appealing. Clearly, because GR has been tested rigorously in the laboratories and solar system \citep{cw2006}, any modifications to it must be strongly suppressed in the local environments and every successful modified gravity model must have some mechanisms to achieve this suppression to pass the first test. 
In this work we will focus on a class of models where GR is modified by an additional dynamical scalar degree of freedom (a scalar field) which mediates a fifth force of gravitational strength between matter particles; the equation of motion of this scalar makes it extremely heavy and therefore hard to propagate, or extremely weakly coupled to matter, in regions of high matter density. In both cases the fifth force is suppressed and therefore much weaker than gravity. The chameleon model of \citep{kw,ms} is a representative example.

Chameleon models can have very rich phenomenology. In many cases, the background evolution can be indistinguishable from that of the standard $\Lambda$CDM paradigm \citep[see, e.g.,][for some examples]{hs2007,lz2009,bbds2008}. The linear perturbations on very large scales are unaffected by the fifth force either, because even in low density environments the range of the fifth force is only of order Mpc \citep{lb2007,lz2009} -- which means that observables such as the cosmic microwave background (CMB) spectrum and integrated Sachs Wolfe (ISW) effect are the same as the $\Lambda$CDM predictions. Finally, because of the strong suppression in the Solar system, there is no detectable deviation from GR locally. Consequently, the only place where we could expect to see effects of modified gravity would be the nonlinear structures such as voids, clusters and galaxies, which are exactly the regime where numerical simulations are needed to make accurate predictions\footnote{Recently, \citet{bv2012} has attempted to use perturbation theory to study the structure formation of modified gravity theories into the nonlinear regime.}.

Although a number of numerical simulations have already been done for such modified gravity models \citep{oyaizu2008,olh2008,sloh2009,lz2009,lz2010,lb2011,zlk2011,lh2011,lztk2012}, they are still at a very early stage. The reason is that the equation governing the scalar field is generally very nonlinear and it usually takes much longer to solve it than 
the standard Poisson equation of Newtonian gravity. Performing 
$N$-body simulations for modified gravity with very large box sizes and high mass/force resolutions are still a technical challenge, and this fact brings us back to the analytical methods, such as the excursion set theory. 

Unfortunately, even the application of the excursion set theory becomes very nontrivial in the modified gravity theories. The nonlinear equation essentially makes the behaviour of the scalar field (and the fifth force) sensitively dependent on the surrounding environment, and to determine the mapping from an initial overdensity to a late-time nonlinear structure we need to have knowledge of the environments, which have very different densities and can at best be described by some probability distributions (see \citet{brs2010} for an earlier study of such mapping). To solve this problem, \citet{le2012} proposed an extension to the standard excursion set theory by solving the above mapping in some specific environments then averaging over the probability distribution of the environments.
An alternative way of viewing this approach is that the critical density for halo formation follows a distribution which depends on the environment density contrast -- in the language of excursion set theory the barrier is stochastic even though the collapse is described by the deterministic spherical collapse model. This stochasticity of barrier is in contrast to the one discussed in \citet{mr10b,ca2011} in the ellipsoidal collapse model and we refer the readers to the Appendix C of \citet{pls2012a} for the discussion regarding the scatter of critical density observed in numerical simulations and the stochasticity of barrier in the excursion set formalism.

The crucial component of the extended excursion set model is the environment: how do we define it? Obviously, different definitions may give rise to different environmental probability distributions, and there must be some physical arguments to motivate the definition. As a first example to illustrate the idea, \citet{le2012} define the environment by fixing  a Lagrangian (or initial comoving) size, which is the simplest possibility, surrounding each proto-halo. As we shall discuss below, this definition has several drawbacks which can be cured by defining the environments as an Eulerian (physical) size.

This paper is organised as follows: in \S~\ref{sect:chameleon} we briefly review the theoretical model to be considered and summarise its main ingredients. We then discuss our new definition of environment and its motivation in \S~\ref{sect:env}. \S~\ref{sect:umf} is the main part of this work, which gives the numerical results of the unconditional mass function using both Lagrangian and Eulerian definitions of environments and discusses their difference; it also compares the two different methods of calculation -- numerical integration and Monte Carlo simulation -- and finds good agreement. The summary and conclusions of this paper can be found in \S~\ref{sect:con}.

\section{The Chameleon Theory}

\label{sect:chameleon}

This section lays down the theoretical framework for investigating the effects of coupled scalar field(s) in cosmology.  We shall present the relevant general field equations in \S~\ref{subsect:equations}, and then specify the models analysed in this paper  in \S~\ref{subsect:specification}.

\subsection{Cosmology with a coupled scalar field}

\label{subsect:equations}

The equations presented in this sub-section can be found in \citep{lz2009, lz2010, lb2011}, and are presented here only to make this work self-contained.

We start from a Lagrangian density
\begin{eqnarray}\label{eq:lagrangian}
{\cal L} =
{1\over{2}}\left[{R\over\kappa}-\nabla^{a}\varphi\nabla_{a}\varphi\right]
+V(\varphi) - C(\varphi){\cal L}_{\rm{DM}} +
{\cal L}_{\rm{S}},
\end{eqnarray}
in which $R$ is the Ricci scalar, $\kappa=8\pi G$ with $G$ being the
gravitational constant, ${\cal L}_{\rm{DM}}$ and
${\cal L}_{\rm{S}}$ are respectively the Lagrangian
densities for dark matter and standard model fields. $\varphi$ is
the scalar field and $V(\varphi)$ its potential; the coupling
function $C(\varphi)$ characterises the coupling between $\varphi$
and dark matter. Given the functional forms for $V(\varphi)$ and $C(\varphi)$
a coupled scalar field model is then fully specified.

Varying the total action with respect to the metric $g_{ab}$, we
obtain the following expression for the total energy momentum
tensor in this model:
\begin{eqnarray}\label{eq:emt}
T_{ab} = \nabla_a\varphi\nabla_b\varphi -
g_{ab}\left[{1\over2}\nabla^{c}\nabla_{c}\varphi-V(\varphi)\right]+ C(\varphi)T^{\rm{DM}}_{ab} + T^{\rm{S}}_{ab},
\end{eqnarray}
where $T^{\rm{DM}}_{ab}$ and $T^{\rm{S}}_{ab}$ are the
energy momentum tensors for (uncoupled) dark matter and standard
model fields. The existence of the scalar field and its coupling
change the form of the energy momentum tensor leading to
potential changes in the background cosmology and 
structure formation.

The coupling to a scalar field produces a direct
interaction (fifth force) between dark matter
particles due to the exchange of scalar quanta. This is best
illustrated by the geodesic equation for dark matter particles
\begin{eqnarray}\label{eq:geodesic}
{{{\rm d}^{2}\bf{r}}\over{{\rm d}t^2}} = -\vec{\nabla}\phi -
{{C_\varphi(\varphi)}\over{C(\varphi)}}\vec{\nabla}\varphi,
\end{eqnarray}
where $\bf{r}$ is the position vector, $t$ the (physical) time, $\phi$
the Newtonian potential and $\vec{\nabla}$ is the spatial
derivative. $C_\varphi\equiv {\rm d}C/{\rm d}\varphi$. The second term in the
right hand side is the fifth force and only exists for coupled matter
species (dark matter in our model). The fifth force also changes the
clustering properties of the dark matter. 

To solve the above two equations we need to know both the time
evolution and the spatial distribution of $\varphi$, {\it i.e.}  we
need the solutions to the scalar field equation of motion (EOM)
\begin{eqnarray}
\nabla^{a}\nabla_a\varphi + {{\rm{d}V(\varphi)}\over{\rm{d}\varphi}} +
\rho_{\rm{DM}}{\frac{\rm{d}C(\varphi)}{\rm{d}\varphi}} = 0,
\end{eqnarray}
or equivalently
\begin{eqnarray}
\nabla^{a}\nabla_a\varphi + {{\rm{d}V_{eff}(\varphi)}\over{\rm{d}\varphi}} =
0,
\end{eqnarray}
where we have defined
\begin{eqnarray}
V_{eff}(\varphi) = V(\varphi) + \rho_{\rm{DM}}C(\varphi).
\end{eqnarray}
The background evolution of $\varphi$ can be solved easily given
the present day value of  $\rho_{\rm{DM}}$ since 
$\rho_{\rm{DM}}\propto a^{-3}$. We can then divide $\varphi$
into two parts, $\varphi=\bar{\varphi}+\delta\varphi$, where
$\bar{\varphi}$ is the background value and $\delta\varphi$ is its
(not necessarily small nor linear) perturbation, and subtract the
background part of the scalar field equation of motion from the full equation
to obtain the equation of motion for $\delta\varphi$. In the
quasi-static limit in which we can neglect time derivatives of
$\delta\varphi$ as compared with its spatial derivatives (which
turns out to be a good approximation on \tcr{galactic and cluster} scales),
we find
\begin{eqnarray}\label{eq:scalar_eom}
\vec{\nabla}^{2}\varphi =
{{\rm{d}C(\varphi)}\over{\rm{d}\varphi}}\rho_{\rm{DM}} -
{{\rm{d}C(\bar{\varphi})}\over{\rm{d}\bar{\varphi}}}\bar{\rho}_{\rm{DM}} +
{{\rm{d}V(\varphi)}\over{\rm{d}\varphi}} -
{{\rm{d}V(\bar{\varphi})}\over{\rm{d}\bar{\varphi}}},
\end{eqnarray}
where $\bar{\rho}_{\rm{DM}}$ is the background dark matter
density.

The computation of the scalar field $\varphi$ using the above equation then completes the computation of the source term for the Poisson equation
\begin{eqnarray}\label{eq:poisson}
\vec{\nabla}^{2}\phi &=& \frac{\kappa}{2}\left[\rho_{\rm tot}+3p_{\rm tot}\right]\nonumber\\
&=& {{\kappa}\over{2}}\left[C(\varphi)\rho_{\rm{DM}} + \rho_{\rm{B}}
- 2V(\varphi)\right], 
\end{eqnarray}
where $\rho_{\rm{B}}$ is the baryon density (we have neglected the kinetic energy of the scalar field because it is always very small for the model studied here).

\subsection{Specification of model}

\label{subsect:specification}

As mentioned above, to fully fix a model we need to specify the functional forms of $V(\varphi)$ and $C(\varphi)$. Here we will use the models investigated by \cite{lz2009, lz2010, li2011}, with
\begin{eqnarray}\label{eq:coupling}
C(\varphi) = \exp(\gamma\sqrt{\kappa}\varphi),
\end{eqnarray}
and 
\begin{eqnarray}\label{eq:pot_chameleon}
V(\varphi) = {{\Lambda}\over{\left[1-\exp\left(-\sqrt{\kappa}\varphi\right)\right]^\alpha}}.
\end{eqnarray}
In the above $\Lambda$ is a parameter of mass dimension four and 
is of order the present dark energy density 
($\varphi$ plays the role of dark energy in the models). $\gamma, \alpha$ are dimensionless  parameters controlling the strength of the 
coupling and the steepness of the potentials respectively. 

We shall choose $\alpha\ll1$ and $\gamma>0$ as in \citet{lz2009, lz2010},
ensuring that $V_{eff}$ has a global minimum close to $\varphi=0$
and ${\rm d}^2V_{eff}(\varphi)/{\rm d}\varphi^2 \equiv m^2_{\varphi}$ at this
minimum is very large in high density regions. There are two
consequences of these choices of model parameters: (1) $\varphi$ is
trapped close to zero throughout  cosmic history so that
$V(\varphi)\sim\Lambda$ behaves as a cosmological constant; 
(2) the fifth force is strongly
suppressed in high density regions where $\varphi$ acquires a large
mass, $m^2_{\varphi}\gg H^2$ ($H$ is the Hubble expansion rate),
and thus the fifth force cannot propagate far. The suppression of the
fifth force is even stronger at early times, and thus its influence on
structure formation occurs mainly at late times.
The environment-dependent behaviour of the scalar field was
first investigated by \citet{kw,ms}, and is often referred
to as the `chameleon effect'.

\section{Discussion on Environment}

\label{sect:env}

The extended excursion set approach for chameleon models proposed in \citet{le2012} differs from the original excursion set approach in the introduction of environment, which is important in the chameleon models for two reasons: first, the environmental density determines the critical density for the spherical collapse inside it; second, an arbitrary spherical overdense region does not reside in an environment with any density equally likely. As a result, in the language of excursion set theory, the calculation of the first-crossing probability of the critical density curve must now be done in different environments and then integrated over the distribution of the environment.

It is therefore evident that the specification of the environment is crucial in the extended excursion set approach. In \citet{le2012}, the environment is defined as the follows:
\begin{enumerate}
\item it is a spherical region with a common centre as the considered spherical overdensity (i.e., the halo-to-be);
\item it is much bigger than the halo-to-be;
\item it is not too big because otherwise it will not give a faithful representation of the environmental density.
\end{enumerate}
As a first approximation, \citet{le2012} defines the environment to have a {\it Lagrangian} radius of $\xi=8h^{-1}$Mpc and calls this a fixed-scale environment approximation. This simplifies the numerical calculation and eliminates the need of Monte Carlo simulations.

However, the use of the Lagrangian radius $\xi$ has certain drawbacks. As a first example, if the environment density is high (e.g., close to the critical density for collapse), then at late times the sizes of the environment and the halo-to-be could be roughly the same, violating the above requirement that the environment should be much bigger than the halo-to-be.  As a second example, if the environmental density is low, then it expands faster than the cosmic expansion and at late times can become very large in size, no longer providing a faithful representation of the environment. 

In \citet{lzk2012} it has been shown, using numerical simulation results, that the analytical formula for the fifth force used in \citet{le2012} is quite accurate, assuming that the environment has an {\it Eulerian} (rather than Lagrangian) radius $\zeta=5\sim8h^{-1}$Mpc; see Fig.~2 there. As a result, a more physically reasonable definition of the environment is 
its Eulerian radius. 
As the characteristic length scale in the chameleon models is 
the Compton length $\lambda_{\mathrm{C}}$ of the scalar field 
(which is a function of time), a physical choice of 
the Eulerian radius $\zeta$ of the environment would then be 
$\zeta\sim\mathcal{O}(\lambda_{\mathrm{C}})$ because matter field 
within $\lambda_{\rm{C}}$ is expected to affect the scalar 
field value at a point (this is the meaning of 'environment').

Of course, $\lambda_{\rm C}$ evolves (increases) in time, while $\zeta$ could either increase or decrease depending on whether the environmental density is lower or higher than the cosmic average. One can certainly choose $\zeta\sim\lambda_{\rm C}$ throughout the evolution, but this will make the computation rather complicated. An alternative is to have $\zeta\sim\lambda_{\rm C}$ at late times, say $z=z_f$ where $z_f$ is the formation redshift of halos. This would mean that at early times $\zeta$ can be different from $\lambda_{\rm C}$, but there are two reasons to expect the overall effect to be small:
\begin{enumerate}
\item the fifth force only becomes important at very late times in most chameleon models, and for redshift $z>2$ it is mostly negligible. At low redshifts the difference between $\zeta$ and $\lambda_{\rm C}$ is small;
\item even at early times, the difference between $\zeta$ and $\lambda_{\rm C}$ is only big for overdense environments which contract during the cosmic history, but for these environments the fifth force is weak throughout the whole cosmic evolution. 
\end{enumerate}
In this work we shall choose $\zeta$ to be smaller than the Lagrangian environment $\xi$ but of the same order as $\lambda_{\rm C}$. 
Because the environment is bigger than 
the range of the fifth force, its evolution could be approximated by the $\Lambda$CDM model. 
The Eulerian overdensity at time $t$ can be related to the linearly 
extrapolated density contrast $\delta(t)$ by 
the spherical collapse model \citep{b1994,sheth98}:
\begin{eqnarray}
\Delta_{\rm NL}(t) = \frac{m}{\bar{\rho}V} \approx \left[1-\frac{\delta(t)}{\delta_{\rm sc}}\right]^{-\delta_{\rm sc}},
\label{eqn:sc}
\end{eqnarray}
where $\delta_{\rm sc}\approx1.676$ is the critical density 
for 
$\Lambda$CDM 
at $z_f=0$, $V$ is the Eulerian volume, $m$ the mass within this Eulerian 
volume and $\bar{\rho}$ is the cosmic background density of matter. 

The distribution of the 
environment density at a given Eulerian scale can be computed by the 
excursion set approach \citep{sheth98,lamsheth08a,pls2012b} using 
the spherical collapse model. 
Effectively Eq.~\eqref{eqn:sc}
defines a curve $B(m)$ in the $\delta$-$S$ plane (see below for  definition of $S$). 
The first crossing of this environment barrier 
gives the value of the linear extrapolated density contrast $\delta(t)$ 
that a spherical region 
containing mass $m$ must have in order to evolve into an Eulerian volume $V$ at $t$:
\begin{eqnarray}
B(m) = \delta_{\rm sc}\left[1-\left(\frac{m}{\bar{\rho}V}\right)^{-1/\delta_{\rm sc}}\right],
\label{eqn:eulerBs}
\end{eqnarray}
where 
the enclosed 
mass is a function of $S$ (ses Eq.~\eqref{eqn:s}). 
The mapping between $m$ and $S$ depends on the linear matter power spectrum 
as well as the smoothing kernel.
If a power-law matter power spectrum $P(k)$ 
with the power index $n_s$ is specified, 
then the above equation can be rewritten as
\begin{eqnarray}\label{eq:B_env}
B(S) = \delta_{\rm sc}\left[1-\left(\frac{\zeta}{8h^{-1}{\rm Mpc}}\right)^{3/\delta_{\rm sc}}\left(\frac{S}{\sigma_8^2}\right)^{\frac{3}{(3+n_s)\delta_{\rm sc}}}\right],
\label{eqn:eulerBspkns}
\end{eqnarray}
where 
\begin{eqnarray}
S(m) = S(\xi) = \frac{1}{2\pi^2}\int^{\infty}_0k^2P(k)W^2(k\xi){\rm d}k,
\label{eqn:s}
\end{eqnarray}
with $W(k\xi)$ the filter function and $\xi$ the Lagrangian radius of the filter so that $m(\xi)=\frac{4}{3}\pi\xi^3\bar{\rho}$. $\sigma_8$ is given by $S=\sigma_8^2$ with $\xi=8h^{-1}$Mpc. 
In particular for $n_s = -1.2$ this barrier becomes linear in $s$ 
(approximating $\delta_{\rm SC} = 5/3$) and 
make the analytical analysis easier.
In what follows we shall use $\zeta=5h^{-1}$Mpc as suggested by \citet{lzk2012}.
One can make a similar analogy for Lagrangian environment, only in 
this case the barrier is vertical (see figure~\ref{fig:env_eu_la})
and there is one and only one crossing of this Lagrangian barrier.

In the language of excursion set theory, then, the first crossing probability of the moving barrier $B(S)$ in $[S,S+dS]$, denoted by $P_{\rm env}(S)dS$, is the probability that an arbitrary point (the centre of a halo-to-be) is in an environment whose linearly extrapolated density contrast falls into $[B(S), B(S+dS)]$ 
and this environment will evolve into a region having an Eulerian radius 
$\zeta$ at $z_f$; the total mass enclosed in this environment is $m(S+dS)\leq M_{\rm env}\leq m(S)$. Both the environmental density and environmental mass are important for the discussions below.

\section{Unconditional Mass Functions}

\label{sect:umf}

As a first application of the idea described above, in the rest of the paper we shall study the (unconditional) mass function of the chameleon models, using both numerical integration and Monte Carlo simulations, assuming
uncorrelated steps in the framework of the excursion set formalism. 
The analysis of correlated steps and the associated 
conditional mass function and halo bias will be discussed in \citet{ll2012}.
 
\subsection{Eulerian versus Lagrangian environments}

\label{subsect:env_eu_la}

\begin{figure}
\includegraphics[width=95mm]{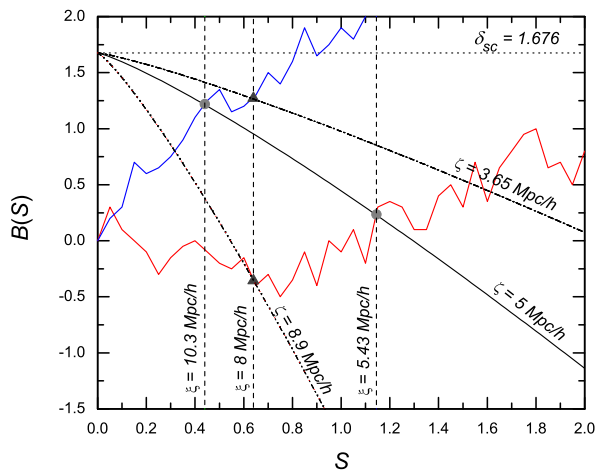}
\caption{(Colour Online) An illustration of the Eulerian versus Lagrangian environments. See the text for a detailed description.}
\label{fig:env_eu_la}
\end{figure}

\begin{figure}
\includegraphics[width=95mm]{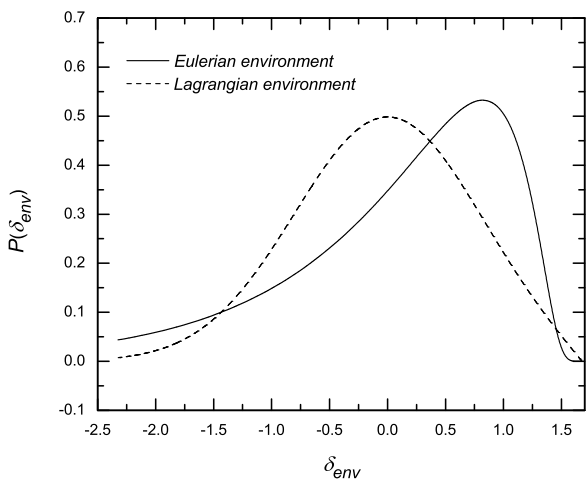}
\caption{The probability distribution, $P\left(\delta_{\rm env}\right)$, of the Eulerian (solid curve) and Lagrangian (dashed curve) environments, normalised to unity.}
\label{fig:env_distr}
\end{figure}

Fig.~\ref{fig:env_eu_la} illustratively demonstrates the difference between
 the two definitions of the environment (Eulerian and Lagrangian). 
Here, the dotted horizontal line denotes the critical density $\delta_{\rm sc}\approx1.676$ for a $\Lambda$CDM collapse, and the middle one of the dashed vertical lines represents the value of $S$ corresponding to a Lagrangian radius $\xi=8h^{-1}$Mpc, which was used in \citet{le2012} as the definition of the (Lagrangian) environment. Consider now two random walk trajectories (the blue and red curves): they cross the middle vertical dashed line at the two dark grey filled triangles, but do not cross the dotted horizontal line before that. Those two triangles correspond to two very different values of $\delta_{\rm env}$ ($1.3$ and $-0.4$) and thus the two random walks represent structure formations in very different (Lagrangian) environments.

The Eulerian environment, on the other hand, is defined such that its Eulerian radius today is $5h^{-1}$Mpc and it is represented by the solid curve. The random walks cross this curve at the two grey filled circles, corresponding to $\delta_{\rm env}\sim1.25$ and $0.25$ respectively: 
the environments are less different if they are Eulerian! In particular, the structure formation relevant for the red random walk now takes place in an overdense rather than underdense environment.

Note that the two filled circles correspond to Lagrangian radii of $\xi=10.3$ and $5.43h^{-1}$Mpc (the left and right vertical dashed lines) respectively, while the two filled triangles correspond to Eulerian radii of $\zeta=3.65$ (dash-dotted curve) and $8.9h^{-1}$Mpc (dash-dot-dot curve) respectively. It is clear that the Eulerian environment of the blue random walk contains about seven times more matter than that of the red random walk, but if one uses Lagrangian definition then both environments contain the same amount of matter.

The above result has implied that the Eulerian and Lagrangian environments should have different probability distribution functions (PDFs). We then need to compute the PDF of $\delta_{\rm env}$, $P(\delta_{\rm env})$, for these environment definitions: the Lagrangian density contrast follows the Gaussian distribution and the corresponding $P(\delta_{\rm env})$ is given by Eq.~(40) of \citet{le2012}, while the Eulerian density can be considered as a mixture of Lagrangian density from different smoothing scales: a dense Eulerian environment is evolved from a bigger Lagrangian batch but the opposite is true for underdense Eulerian environments.

There are different methods to calculate $P(\delta_{\rm env})$ for the Eulerian environment: one can either obtain it by solving 
the first crossing distribution of the barrier in Eq.~\eqref{eqn:eulerBs} (see \cite{zh2006,lamsheth09} for uncorrelated steps; \cite{mr10,ca2011,pls2012a,ms12} for correlated steps), or 
using analytical expressions such as  the log-normal distribution or the expressions given in \cite{lamsheth08a,lamsheth08b} for the evolved nonlinear density contrast and combining that with 
Eq.~\eqref{eqn:sc} 
to obtain $P(\delta_{\rm env})$ (recall that $\delta_{\rm env}$ is the linearly extrapolated density contrast at the Eulerian environment).
In particular, for power-law matter power spectrum the barrier is given by Eq.~\eqref{eqn:eulerBspkns} and applying the first crossing probability approximation in \citet{lamsheth08a} (see \citet{lamsheth09} for an explanation for this approximation), the 
distribution of $\delta_{\rm env}$ is
\begin{align}
P(\delta_{\rm env}) = & \frac{\beta^{\omega/2}}{\sqrt{2\pi}}
     \left[1 + \left(\omega-1\right)\frac{\delta_{\rm env}}{\delta_c}\right]
 \left(1-\frac{\delta_{\rm env}}{\delta_c}\right)^{-\omega/2-1} \nonumber \\*
&\times      \exp\left[-\frac{\beta^{\omega}}{2}\frac{\delta_{\rm env}}{(1-\delta_{\rm env}/\delta_c)^\omega}\right],
\end{align}
where $\beta = (\zeta/8)^{3/\delta_c}/\sigma_8^{2/\omega}$, 
$\omega = \delta_c\gamma$, and $\gamma$ is the logarithmic derivative 
of the density fluctuation variance w.r.t. $m$:
\begin{equation}
\gamma = -\frac{{\rm d}\ln S}{{\rm d}\ln m} = \frac{n_s+3}{3}.
\end{equation}
The case where $n_s=-1.2$ is of special interest, 
since not only 
the barrier $B(S)$ is linear in $S$ (if one set 
$\delta_c = 5/3$),  
the first crossing probability approximation above is also exact 
 \citep[see, for example,][for derivation]{lamsheth09}. 
The associated distribution of $\delta_{\rm env}$ becomes
\begin{equation}
 P(\delta_{\rm env}) = \frac{1}{\sqrt{2\pi}}\frac{\beta^{1/2}}{(1-\delta_{\rm env}/\delta_c)^{3/2}} e^{-\delta_{\rm env}^2\beta/2(1-\delta_{\rm env}/\delta_c)}.
\end{equation}

The two environment density distributions for the case of power-law power 
spectrum with $n_s=-1.5$ are shown in Fig.~\ref{fig:env_distr}, 
which shows the PDF for the Eulerian environment peaks at bigger $\delta_{\rm env}$ than that for the Lagrangian environment. 
The results were computed numerically using the method described in \citet{zh2006} and used in \citet{le2012}. Note that
\begin{enumerate}
\item If $\delta_{\rm env}$ is close to $\delta_{\rm sc}$ the PDF is smaller for the Eulerian environment, because for such an environment to evolve into an Eulerian radius of $5h^{-1}$Mpc today, it must have an extremely large Lagrangian size ($\xi\gg8h^{-1}$Mpc), which is a very rare event;
Note that the other reason is due to the approximation formula Eq.~\eqref{eqn:sc} -- it maps $\delta_{\rm lin}\rightarrow \delta_c$ to $\delta_{\rm nl}\rightarrow \infty$. Hence $p(\delta_{\rm env})$ must go to zero for $\delta_{\rm env} \geq \delta_c$.
\item The PDF for Lagrangian environment peaks at $\delta_{\rm env}\approx0$, because at any value of $S$ (say $S=0.64$ for $\xi=8h^{-1}$Mpc, middle vertical dashed line of Fig.~\ref{fig:env_eu_la}) the value of the random walk position satisfies a Gaussian distribution;
\item The PDF for Eulerian environment peaks at $\delta_{\rm env}\approx0.8>0$, because the random walk is crossing a decreasing barrier. This means that if one integrates over the PDF of environments then more contributions come from higher-density environments under the Eulerian definition;
\item The PDF is lower for small $\delta_{\rm env}$ ($<-1.5$) under the Lagrangian definition, because underdensities with linearly extrapolated density contrast $\delta<-1.5$ and Lagrangian size $\xi=8h^{-1}$Mpc will have evolved into very large Eulerian sizes today ($\zeta\gg5h^{-1}$Mpc), and this is rare event; in contrast, to have an Eulerian size of $5h^{-1}$Mpc today such undensities can be quite small in initial sizes whose r.m.s fluctuation is bigger and hence 
the probability is not negligible.
\end{enumerate}

\subsection{Spherical collapse in a given environment}

\label{subsect:collapse}

\begin{figure}
\includegraphics[width=95mm]{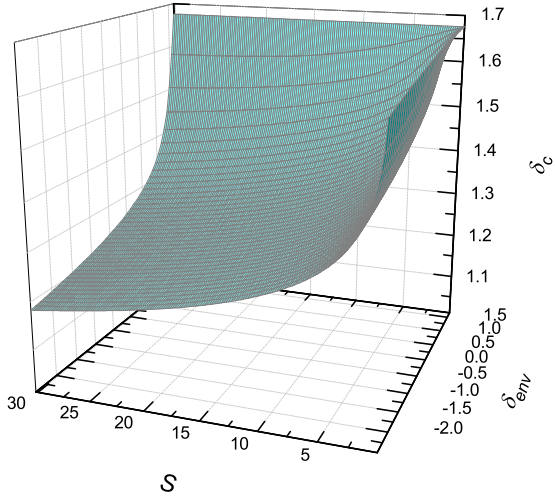}
\caption{The critical density for the spherical collapse at $z_f=0$, as a function of the linearly extrapolated environmental density contrast $\delta_{\rm env}$ and $S$.}
\label{fig:dc_grid}
\end{figure}

\begin{figure*}
\includegraphics[width=175mm]{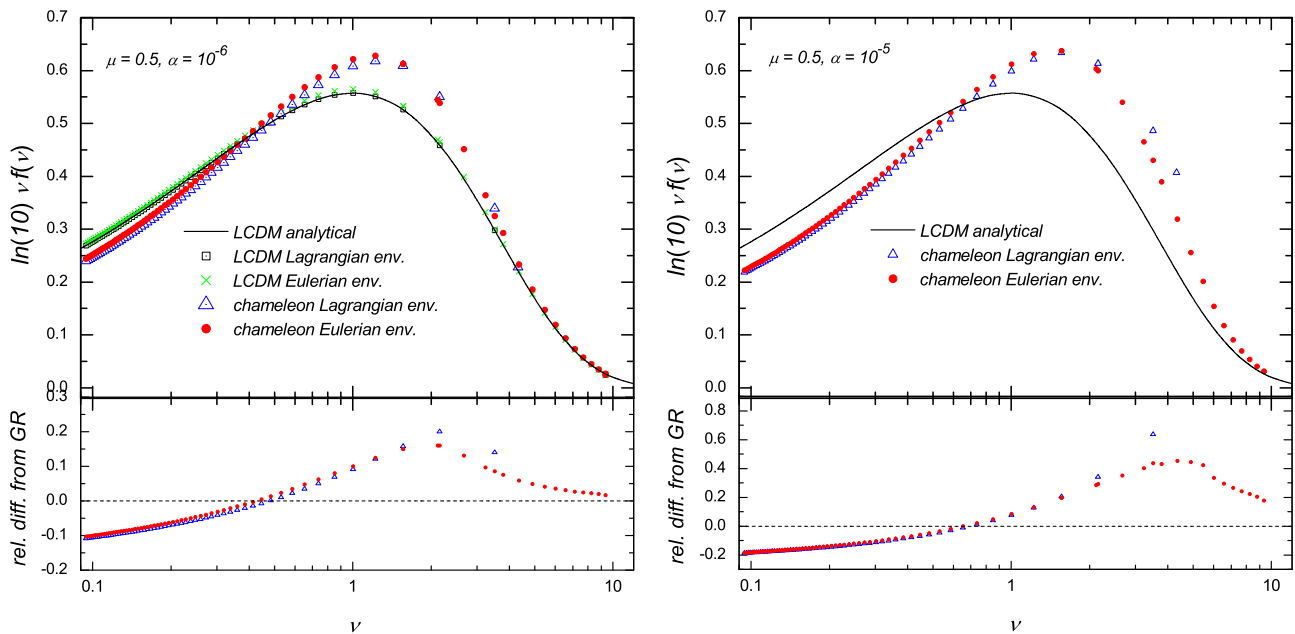}
\caption{(Colour Online) {\it Upper left panel}: the collapsed mass fraction $\nu F(\nu)$ for the $\Lambda$CDM and chameleon model with $\alpha=10^{-6}, \mu=0.5$; for each model the environmental average (see the text) is performed assuming Lagrangian and Eulerian environments respectively (see the legends), and the solid curve is the analytic result for $\Lambda$CDM. {\it Upper right panel}: the same as the upper left panel, but here only the results for a chameleon model with $\alpha=10^{-5}, \mu=0.5$ (the symbols) are shown; the analytical results for $\Lambda$CDM is shown as a solid curve for comparison. {\it Lower left panel}: the relative difference of $\nu F(\nu)$ between the chameleon model with $\alpha=10^{-6}$ and $\Lambda$CDM. {\it Lower right panel}: the relative difference of $\nu F(\nu)$ between the chameleon model with $\alpha=10^{-5}$ and $\Lambda$CDM. Note that the vertical axis is $\ln(10)\nu F(\nu)$ rather than $\nu F(\nu)$.}
\label{fig:nufnu_ni}
\end{figure*}

The spherical collapse history of an initial overdensity in a given environment of linearly extrapolated density contrast $\delta_{\rm env}$ has been discussed in detail in \citet{le2012}, to
which 
interested readers are referred, 
and here we will only give a qualitative description of the results.

Fig.~\ref{fig:dc_grid} shows the critical density $\delta_{\rm c}(S,\delta_{\rm env})$ for an initial overdensity to collapse at $z_f=0$, as a function of both $S$ and $\delta_{\rm env}$. As the fifth force is always attractive, $\delta_{\rm c}(S,\delta_{\rm env})<\delta_{\rm sc}$ where $\delta_{\rm sc}$ is the critical density if the fifth force vanishes. 
Furthermore,
\begin{enumerate}
\item in high-density environments ($\delta_{\rm env}\rightarrow\delta_{\rm sc}$) the fifth force is strongly suppressed so that the collapse is governed by Newtonian gravity only and $\delta_{\rm c}\approx\delta_{\rm sc}$;
\item as $\delta_{\rm env}$ decreases, the fifth force becomes stronger in general and enhances the matter clustering. This means that 
the required critical density to collapse at $z_f$ is lower;
\item very small values of $S$ correspond to the very big smoothing scales.
Such regions 
have efficient self-screening of the fifth force 
irrespective of their environment, and hence $\delta_{\rm c}(S,\delta_{\rm env})\approx\delta_{\rm sc}$ for arbitrary $\delta_{\rm env}$ for small $S$;

\item large values of $S$ correspond to small smoothing scales, 
which does not have strong self-screening of the fifth force, and 
hence $\delta_{\rm c}(S, \delta_{\rm env})$ depends sensitively on $\delta_{\rm env}$.
\end{enumerate}

In practice, once we know $\delta_{\rm env}$, we can fully determine the collapse 
criteria
for halos of arbitrary size in this environment, and 
therefore the first crossing probability density, $f(S|\delta_{\rm env})$,
 across the barrier $\delta_{\rm c}(S,\delta_{\rm env})$.
Here $|$ denotes it is the conditional first crossing probability 
that 
the random walks having \textit{first} crossed $B(S)$ given in Eq.~(\ref{eq:B_env}) at $B(S)=\delta_{\rm env}$. 
One then only needs to average over the first crossing 
probability distribution of $\delta_{\rm env}$ 
(which we have found in the previous subsection) to 
find the averaged first crossing probability $f(S)$.

\subsection{Averaging over environmental distribution}

\label{subsect:average}

As in \citet{le2012}, the final first crossing probability, which is related to the unconditional mass function, is calculated by making the environmental average:
\begin{eqnarray}\label{eq:num_int}
f(S) &=& \int_{-\infty}^{\delta_{\rm sc}}f(S | \delta_{\rm env})P(\delta_{\rm env}){\rm d}\delta_{\rm env},
\end{eqnarray} 
where the upper limit of the integral is $\delta_{\rm sc}$ because $\delta_{\rm env}\leq\delta_{\rm sc}$ (c.f.~Fig.~\ref{fig:env_distr}) because by definition an environment has not collapsed to form a halo by $z_f$. In the case of Lagrangian environment, $P({\delta_{\rm env}})$ is defined so that it is identically zero for $\delta_{\rm env}\geq\delta_{\rm sc}$, as shown in Eq.~(40) of \citet{le2012}. For the Eulerian environment proposed in this work, the Eulerian barrier always lies below $\delta_{\rm sc}$ for all $S>0$ hence this upper limit is valid by construction.

The mass function, $dn/dM$, is related to $f(S)$ by
\begin{eqnarray}
\frac{{\rm d}n}{{\rm d}M}{\rm d}M &=& \frac{\bar{\rho}_m}{M}f(S)\left|\frac{{\rm d}S}{{\rm d}M}\right|{\rm d}M,
\end{eqnarray}
where $n$ is the number density of halos and $M$ the halo mass. In the literature, an alternative quantity which is often used is the collapsed mass fraction $\nu F(\nu)$, with $\nu\equiv\delta^2_{\rm sc}/S$. $\nu F(\nu)$ is the fraction of matter in collapsed objects per logarithmic interval of $\nu$ and satisfies
\begin{eqnarray}
F(\nu){\rm d}\nu\ = \ f(S){\rm d}S &\Rightarrow& \nu F(\nu)\ =\ Sf(S). 
\end{eqnarray}

In the following subsections, we shall calculate the 
quantity $\nu F(\nu)$ using two methods: the numerical integration of Eq.~(\ref{eq:num_int}) and Monte Carlo simulations. 
We focus on the excursion set with uncorrelated step in the current study 
but would like to point out that Eq.~(\ref{eq:num_int})
applies in both correlated and uncorrelated steps calculation --
in the case of correlated steps one must take into account the fact that 
the condition in $f(S|\delta_{\rm env})$ is the random walk \textit{first} crosses the environment barrier at 
$\delta_{\rm env}$, or in other words the random walk is non-Markovian.
It complicates the analytical calculation and 
it will be discussed in \citet{ll2012}. 
In the present case where the random walk has uncorrelated steps 
(it is Markovian), the conditional probability does not depend on 
the history of the walk prior to $\delta_{\rm env}$.

In what follows we apply two approaches to evaluate the halo mass function in modified gravity models. The numerical integration approach is similar to the previous work by \citet{le2012} which applies the method in \citet{zh2006}; the Monte Carlo simulation approach follows the variation of density contrast as a function of smoothing scale and we keep tracks of the first crossing of the environment barrier as well as the first crossing of the consequent modified halo formation barrier.

\subsubsection{Numerical integration}

\label{subsubsect:ni}

As numerical examples, we have calculated $\nu F(\nu)$ as a function of $\nu$ for three models -- GR\footnote{We will consider GR with a cosmological constant which drives the accelerating expansion of the Universe. So in what follows we will use the words 'GR' and '$\Lambda$CDM' interchangeably.} and two chameleon models with $\mu=0.5, \alpha=10^{-6}$ and $\mu=0.5, \alpha=10^{-5}$ -- all having the same background cosmology.  A power-law matter power spectrum with index $n_s=-1.5$ is assumed. For each model we have done the calculation assuming Lagrangian and Eulerian environments respectively, and the results are shown in Fig.~\ref{fig:nufnu_ni}.

Assuming spherical collapse, there is no environmental dependence in the $\Lambda$CDM result, which means that the prediction of $\nu F(\nu)$ should be the same whether one uses the Lagrangian or Eulerian definition of environment. This has been confirmed in the upper left panel of Fig.~\ref{fig:nufnu_ni}, which serves as a check of accuracy of the numerical computation. For comparison we have also plotted the exact analytic solution for the $\Lambda$CDM model (solid curve)
\begin{eqnarray}
F(\nu) = \sqrt{\frac{\nu}{2\pi}}\exp\left(-\frac{\nu}{2}\right).
\end{eqnarray}

In Fig.~\ref{fig:env_distr} we have seen that the PDF for Eulerian environment peaks at higher $\delta_{\rm env}$ than the PDF for Lagrangian environment. Because higher environmental density means stronger  suppression and therefore weaker effect of the fifth force, we would expect that the deviation from $\Lambda$CDM prediction of $\nu F(\nu)$ be smaller if we use the Eulerian instead of Lagrangian definition of environment. This is confirmed by Fig.~\ref{fig:nufnu_ni}, which shows that the former gives smaller $\nu F(\nu)$ for medium and large $\nu$ (halos of medium and large sizes).

For small values of $\nu$, on the other hand, $\nu F(\nu)$ is bigger if one uses Eulerian environment, because a smaller fraction of matter has fallen into medium and large collapsed halos compared to the case of Lagrangian environments, and therefore more matter is left to form the isolated small halos.

Note that one disadvantage of using a (fixed-sized) Lagrangian radius, as in \citet{le2012}, is that all the environments have exactly the same initial size and so contain the same amount of matter. In particular, the biggest halo to form in any given environment cannot contain more matter than that is contained in the environments, and this places an upper limit of the halo mass, as is shown in Fig.~\ref{fig:nufnu_ni}. If one uses Eulerian environment this problem is solved because the environment can be arbitrarily large in size.

\subsubsection{Monte Carlo simulations}

\label{subsubsect:mc}

\begin{figure}
\includegraphics[width=95mm]{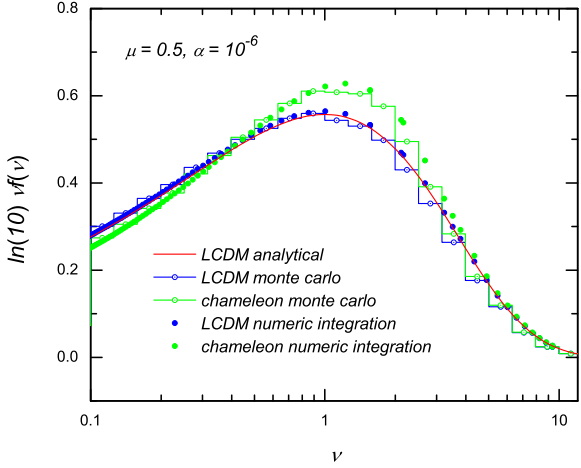}
\caption{(Colour Online) Comparison of the mass fraction functions obtained from numerical integration above and Monte Carlo simulations. See the legends for the details.}
\label{fig:nufnu_ni_mc}
\end{figure}

\begin{figure*}
\includegraphics[width=0.8\textwidth]{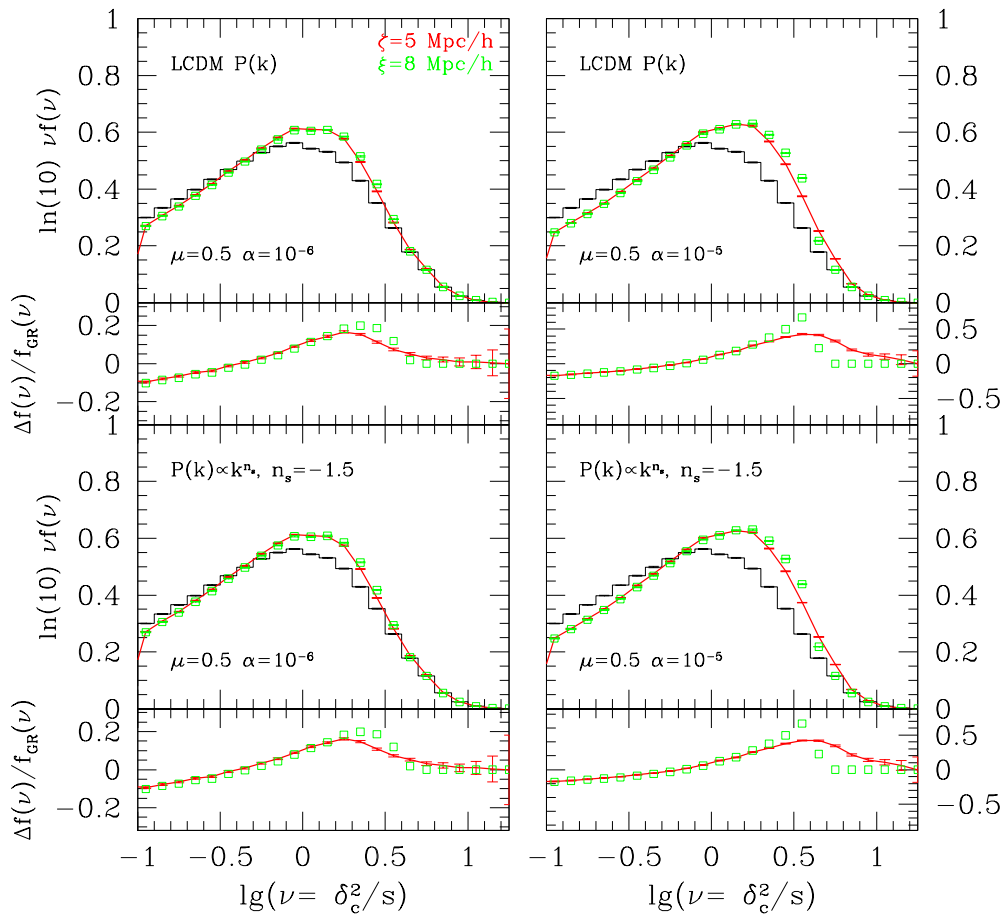}
\caption{(Colour Online) Comparison of mass function obtained from Monte Carlo ssimulation. {\it Left column}: chameleon model with $\mu=0.5,\alpha=10^{-6}$; 
{\it right column}: $\mu=0.5,\alpha=10^{-5}$. The upper row shows results
using $\Lambda$CDM power spectrum while the lower row shows results with 
power-law power spectrum. In each panel the black histogram shows the 
GR + $\Lambda$CDM mass function. Two choices of environment are chosen for the 
chameleon models: Eulerian with $\zeta=5h^{-1}$Mpc (red solid curves), Lagrangian with $\xi=8h^{-1}$Mpc (green boxes). Only error bars for the red solid curves are shown in the lower panel for clarity.}
\label{fig:nufnu_mc}
\end{figure*}

\begin{figure}
\includegraphics[width=0.5\textwidth]{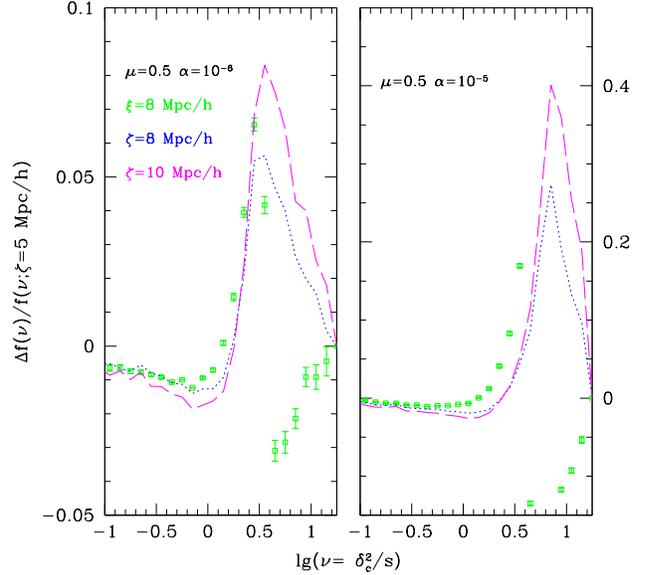}
\caption{(Colour Online) Comparison of the effect of mass function for different environment definitions in the chameleon models. Results are obtained 
from Monte Carlo simulations with $\Lambda$CDM power spectrum where 
ratios of mass function are taken with respect to that of Eulerian environment 
$\zeta=5 h^{-1}$Mpc (red solid curve in figure~\ref{fig:nufnu_mc}). 
Three other chameleon models are considered: 
Lagrangian environment with $\xi=8h^{-1}$Mpc (green boxes); 
Eulerian environment with $\zeta=8h^{-1}$Mpc (blue dotted curves) and with 
$\zeta=10h^{-1}$Mpc (magenta dashed curves).
{\it Left panel}: chameleon model with $\mu=0.5,\alpha=10^{-6}$; 
{\it right panel}: $\mu=0.5,\alpha=10^{-5}$. 
Only errors bars for Lagrangian environment with $\xi=8h^{-1}$Mpc (green boxes)
are shown for clarity.}
\label{fig:nufnu_ratio}
\end{figure}

In this subsection we compare the first crossing distribution 
for different models with or without the chameleon fifth force using Monte Carlo
simulations. 
It is customary to use the variance in 
Eq.~\eqref{eqn:s} with a tophat window function to relate 
$S$ and the smoothing scale $R$ although strictly 
speaking it is not fully consistent since 
random walks with the tophat window function induces correlated steps (or 
one needs to use the sharp $k$-space window function -- which does not 
have a well defined enclosed mass -- to 
obtain uncorrelated steps) \citep{bcek}.

In the Monte Carlo simulations a sample of random walks is generated 
following the procedure in \citet{bcek}. In hierarchical models, 
the variance of the density field is a monotonic 
function of smoothing scale. Hence variance of the density field $S$, 
the smoothing scale as well as the total mass enclosed are interchangeable 
quantities.
For the case of $\Lambda$CDM a constant 
barrier with $\delta_{\rm sc} = 1.676$ is assumed. 
For chameleon models the environment density $\delta_{\rm env}$ is recorded
at the corresponding scales:
for Lagrangian environments, the environment
density $\delta_{\rm env,L}$ is recored at $\xi = 8h^{-1}$Mpc;
for Eulerian environment $\delta_{\rm env,E}$ is the height of the random walk 
where it first crosses the barrier $B(S)$  in Eq.~\eqref{eqn:eulerBs}.
The halo formation criterion $\delta_c(S,\delta_{\rm env})$ 
is determined using this environment density
$\delta_{\rm env}$ (see Fig.~\ref{fig:dc_grid}).
We then follow the random walk until it first crosses 
$\delta_c(S,\delta_{\rm env})$ and record the associated value of $S$.

Note that an advantage of using an Eulerian scale as environment is that the Eulerian barrier $B(S)$ by definition always lies below the constant barrier -- hence it is impossible to reach $\delta_{\rm sc}$ before crossing $B(S)$ and an environment always contains more matter than the halo inside it does. On the other hand, if one uses Lagrangian environments, it is possible, though rare, that the random walk reaches 
$\delta_{\rm sc}$ at a Lagrangian scale bigger than the predefined Lagrangian environment radius $\xi$: this is just the problem we discussed at the end of Sect.~\ref{subsubsect:ni}, and in this case we assume that the evolution of this huge overdensity follows the background environment ($\delta_{\rm env}=0$). For $\xi = 8h^{-1}$Mpc it corresponds to first crossing across $\delta_{\rm sc}$ 
beyond $\lg(\nu) \approx 0.63$.

Fig.~\ref{fig:nufnu_ni_mc} compares the output of the Monte Carlo simulations to that from the numerical integration described in the previous subsection. 
For this purpose we have only plotted the $\Lambda$CDM model and a chameleon model of $\mu=0.5, \alpha=10^{-6}$, adopting the Eulerian definition of the environment with a radius of $5h^{-1}$Mpc. 
The predicted mass functions from the two methods agree with each other 
reasonably in both models and, in the former case, they 
are consistent with the analytical result as well.

As Monte Carlo simulations with uncorrelated steps are generally much faster than the numerical integration, we shall use the former to test the effects of 
different physical parameters. Fig.~\ref{fig:nufnu_mc} and \ref{fig:nufnu_ratio} show the dependency of the collapsed mass fraction $\nu F(\nu)$ on the definition of environment, the initial power spectrum and the model parameter $\alpha$, from which we can see that
\begin{enumerate}
\item increasing the radius of the Eulerian environment results in 
       more big halos because it necessarily means that the PDF of 
       the initial environment density $\delta_{\rm env}$ shifts towards 
       lower values of $\delta_{\rm env}$, 
        making the formation of such halos more strongly 
       affected by the fifth force. 
This dependence on the Eulerian environment radius is subdominant compared to the modification to the GR case;
\item the effect of the Eulerian environmental radius on the abundance of small halos is weaker;
\item switching to a Lagragian environment results in a distinctive  drop in intermediate mass halo -- this decrease corresponds to the Lagrangian scale chosen: beyond $\lg(\nu) = 0.63$ our Lagrangian  environment chameleon models reverts to the GR mass function;
\item the results using a $\Lambda$CDM matter power spectrum are similar to the power-law power spectrum with $n_s=-1.5$;
\item increasing $\alpha$ weakens the chameleon effect which suppresses the fifth force in high matter-density regions, and thus producing more large halos.
\end{enumerate}

\section{Discussions and Conclusions}

\label{sect:con}

For modified gravity models which reduce to Newton gravity in high density environments such as the Solar system, while at the same time have significant deviations from GR on Mpc scales, the environment is often an important concept 
in both the theory and the techniques used to analyse it.
In this paper we extended the work of \citet{le2012} on estimating
the modification of halo abundance in chameleon models 
within the framework of the excursion set approach by considering
the possibility of defining environments according to their Eulerian, rather than Lagrangian, sizes. 
Being the physical size, the Eulerian size changes with time
and describes the dynamical nature of the true environment. 
By choosing the Eulerian size to be of the same order as the Compton length of the scalar field, the single length scale in the theory, we have a better motivated definition of the environments. 

Of course, the exact value of the Eulerian size $\zeta$ is still a 
free parameter which can be tuned to match the simulation data. 
Alternatively, one can make certain approximations in the calculation 
to obtain the functional form 
of the mass function and then calibrate the parameters using simulations.
Another possible solution is by applying correlated steps in the excursion 
set approach (see \citet{ll2012}) -- dramatic fluctuations between  
similar smoothing scales are unlikely with correlated steps and 
may be able to evade this ambiguity in the choice of Eulerian size.

The Lagrangian definition of environments also suffers from a limitation, namely that all environments, having the same initial comoving size, contain the same amount of matter. 
This effectively sets an upper limit of the halo mass that can be studied.
The Eulerian definition of environment solves this problem because here in principle the environment can be infinitely large, 
and halos of any mass can form within it.

In our theoretical framework, the effect of
 different definitions of the environment enters into the calculation 
through the change in the PDF of $\delta_{\rm env}$, 
which describes the distribution of density contrast in the environment
surrounding halo.
Since the strength of the fifth force depends on $\delta_{\rm env}$, 
modifications in $p(\delta_{\rm env})$ alters the halo abundance for
different environment defintions. 
In particular we found that it is more likely to have high values of 
$\delta_{\rm env}$ in Eulerian environment. 
In this case the fifth force is suppressed and the net result is 
smaller deviation of the mass function from $\Lambda$CDM predictions, 
especially for the big and medium-sized halos.
We have verified this using two methods: numerical integration 
and Monte Carlo simulations, and both methods agree with each other.

To briefly summarise, this work emphasises the importance of environment definition in the study of structure formation in modified gravity theories, and 
lays down the formalism for applying a physically motiviated Eulerian environment in the framework 
of the excursion set approach. This framework could easily be generated to other types of modified gravity theories, such as the envrionmentally dependent dilation \citep{bbds2010,bbdls2011} and the symmetron \citep{hk2010,dlmw2011} modes. Within this framework we can easily analyse other quantities of interests, 
such as the halo bias, voids statistics and merger tree. 
We can also derive approximate analytical expression 
for the halo mass function and calibrate the parameters 
with the numerical simulations. These will be the topics of future works.

\

\section*{Acknowledgments}

BL is supported by the Royal Astronomical Society and the Department of Physics of Durham University, and acknowledges the host of IPMU where this work was initiated. TYL is supported in part by Grant-in-Aid for Young
Scientists (22740149) and by WPI Initiative, MEXT, Japan.

%

\label{lastpage}


\begin{thebibliography}{}
\bibitem[\protect\citeauthoryear{Bertschinger}{1998}]{bertschinger1998} Bertschinger E., 1998, Ann.~Rev.~Astron.~Astrophys., 36, 599
\bibitem[\protect\citeauthoryear{Bernardeau}{1994}]{b1994} Bernardeau F., 1994, ApJ, 427, 51
\bibitem[\protect\citeauthoryear{Bond~{et~al}.}{1991}]{bcek} Bond J.~R., Cole S., Efstathiou G., Kaiser N., 1991, ApJ, 379, 440
\bibitem[\protect\citeauthoryear{Brax, Rosenfeld \& Steer}{2008}]{brs2010} Brax P., Rosenfeld R., Steer, D.~A., 2010, JCAP, 08, 033
\bibitem[\protect\citeauthoryear{Brax~{et~al}.}{2008}]{bbds2008} Brax P., van de Bruck C., Davis A.~C., Shaw D.~J., 2008, PRD, 78, 104021
\bibitem[\protect\citeauthoryear{Brax~{et~al}.}{2010}]{bbds2010} Brax P., van de Bruck C., Davis A.~C., Shaw D.~J., 2010, PRD, 82, 063519
\bibitem[\protect\citeauthoryear{Brax~{et~al}.}{2011}]{bbdls2011} Brax P., van de Bruck C., Davis A.~C., Li B., Shaw D.~J., 2011, PRD, in press
\bibitem[\protect\citeauthoryear{Brax \& Valageas}{2012}]{bv2012} Brax P., Valageas P., 2012, arXiv:1205.6583
\bibitem[\protect\citeauthoryear{Corasaniti \& Achitouv}{2011}]{ca2011} Corasaniti P.~S., Achitouv I., 2011, PRD, 84, 23009
\bibitem[\protect\citeauthoryear{Copeland~{et~al}.}{2006}]{cst2006} Copeland E.~J., Sami M., Tsujikawa S., 2006, IJMPD, 15, 1753
\bibitem[\protect\citeauthoryear{Clifton~{et~al}.}{2011}]{cfps2011} Clifton T., Ferreira P.~G., Padilla A., Skordis C., 2011, arXiv:1106.2476 [astro-ph.CO]
\bibitem[\protect\citeauthoryear{Davis {et~al.}}{2011}]{dlmw2011} Davis A.~C., Li B., Mota D.~F., Winther H.~A., 2011, ApJ., in press; arXiv:1108.3082 [astro-ph.CO]
\bibitem[\protect\citeauthoryear{Hinterbichler \& Khoury}{2010}]{hk2010} Hinterbichler K., Khoury J., 2010, PRL, 104, 231301
\bibitem[\protect\citeauthoryear{Hu \& Sawicki}{2007}]{hs2007} Hu W.,
  Sawicki I., 2007, PRD, 76, 064004
\bibitem[\protect\citeauthoryear{Jain \& Khoury}{2010}]{jk2010} Jain B.,
  Khoury J., 2010, Annals Phys., 325, 1479
\bibitem[\protect\citeauthoryear{Jenkins~et~al.}{2001}]{jetal2001} Jenkins A.~R., Frenk C.~S., White S.~D.~M., Colberg J.~M., Cole S., Evrard A.~E., Couchman H.~M.~P., Yoshida N., 2001, MNRAS, 321, 372
\bibitem[\protect\citeauthoryear{Khoury \& Weltman}{2004}]{kw} Khoury J., Weltman A., 2004, PRD, 69, 044026
\bibitem[\protect\citeauthoryear{Lam \& Sheth}{2008a}]{lamsheth08a} Lam T. Y., Sheth R. K., 2008a, MNRAS, 386, 407
\bibitem[\protect\citeauthoryear{Lam \& Sheth}{2008b}]{lamsheth08b} Lam T. Y., Sheth R. K., 2008b, MNRAS, 389, 1249
\bibitem[\protect\citeauthoryear{Lam \& Sheth}{2009}]{lamsheth09} Lam T. Y., Sheth R. K., 2009, MNRAS, 398, 2143
\bibitem[\protect\citeauthoryear{Lam \& Li}{2012}]{ll2012} Lam T. Y., Li B., 2012, MNRAS, submitted; arXiv:1205.0059 [astro-ph.CO]
\bibitem[\protect\citeauthoryear{Li}{2011}]{li2011} Li B., 2011, MNRAS, 411, 2615
\bibitem[\protect\citeauthoryear{Li \& Barrow}{2007}]{lb2007} Li B., Barrow J.~D., 2007, PRD, 75, 084010
\bibitem[\protect\citeauthoryear{Li \& Barrow}{2011}]{lb2011} Li B., Barrow J.~D., 2011, PRD, 83, 024007
\bibitem[\protect\citeauthoryear{Li \& Efstathiou}{2012}]{le2012} Li B., Efstathiou G., MNRAS, in press; arXiv:1110.6440 [astro-ph.CO]
\bibitem[\protect\citeauthoryear{Li, Zhao \& Koyama}{2012}]{lzk2012} Li B., Zhao G., Koyama K., 2012, MNRAS, in press; arXiv:1111.2602 [astro-ph.CO]
\bibitem[\protect\citeauthoryear{Li, Zhao, Teyssier \& Koyama}{2012}]{lztk2012} Li B., Zhao G., Teyssier R., Koyama K., 2012, JCAP, 1201, 051
\bibitem[\protect\citeauthoryear{Li \& Zhao}{2009}]{lz2009} Li B., Zhao H., 2009, PRD, 80, 044027
\bibitem[\protect\citeauthoryear{Li \& Zhao}{2010}]{lz2010} Li B., Zhao H., 2010, PRD, 81, 104047
\bibitem[\protect\citeauthoryear{Li \& Hu}{2011}]{lh2011} Li Y., Hu W., 2011, PRD, 84, 084033
\bibitem[\protect\citeauthoryear{Ma~{et~al.}}{2011}]{maetal11} Ma C., Maggiore M., Riotto A., Zhang J., 2011, MNRAS, 411, 2644
\bibitem[\protect\citeauthoryear{Maggiore \& Riotto}{2010}]{mr10} Maggiore M., Riotto A., 2010, ApJ, 711, 907
\bibitem[\protect\citeauthoryear{Maggiore \& Riotto}{2010}]{mr10b} Maggiore M., Riotto A., 2010, ApJ, 711, 515
\bibitem[\protect\citeauthoryear{Mota \& Shaw}{2007}]{ms} Mota D.~F., Shaw, D.~J., 2007, PRD, 75, 063501
\bibitem[\protect\citeauthoryear{Musso \& Sheth}{2012}]{ms12} Musso M., Sheth R. K., 2012, arXiv:1201.3876 [astro-ph.CO]
\bibitem[\protect\citeauthoryear{Oyaizu}{2008}]{oyaizu2008} Oyaizu H., 2008, PRD, 78, 123523
\bibitem[\protect\citeauthoryear{Oyaizu~{et~al.}}{2008}]{olh2008} Oyaizu H., Lima M., Hu W., 2008, PRD, 78, 123524
\bibitem[\protect\citeauthoryear{Paranjape \& Sheth}{2012}]{ps2012} Paranjape A., Sheth R. K., 2012, MNRAS, 419, 132
\bibitem[\protect\citeauthoryear{Paranjape, Lam \& Sheth}{2012a}]{pls2012a} Paranjape A., Lam T. Y., Sheth R. K., 2012, MNRAS, 420, 1429
\bibitem[\protect\citeauthoryear{Paranjape, Lam \& Sheth}{2012b}]{pls2012b} Paranjape A., Lam T. Y., Sheth R. K., 2012, MNRAS, 420, 1648
\bibitem[\protect\citeauthoryear{Perlmutter~{et~al.}}{1999}]{perlmutter1999} Perlmutter S.~{\it et.~al.}, 1999, ApJ, 517, 565
\bibitem[\protect\citeauthoryear{Riess~{et~al.}}{1998}]{riess1998} Riess A.~G.~{\it et.~al.}, 1998, Astron.~J., 116, 1009
\bibitem[\protect\citeauthoryear{Schmidt~{et~al.}}{2009}]{sloh2009} Schmidt F., Lima M., Oyaizu H., Hu W., 2009, PRD, 79, 083518
\bibitem[\protect\citeauthoryear{Sheth}{1998}]{sheth98} Sheth R.~K., 1998, MNRAS, 300, 1057
\bibitem[\protect\citeauthoryear{Sheth \& Tormen}{1999}]{st1999} Sheth R.~K., Tormen G., 2002, MNRAS, 308, 119
\bibitem[\protect\citeauthoryear{Sheth \& Tormen}{2002}]{st2002} Sheth R.~K., Tormen G., 2002, MNRAS, 329, 61
\bibitem[\protect\citeauthoryear{Tinker~et~al.}{2008}]{tetal2008} Tinker J.~L., Kravtsov A.~V., Klypin A., Abazajian K., Warren M.~S., Yepes G., Gottlober S., Holz D.~E., 2008, ApJ, 688, 709
\bibitem[\protect\citeauthoryear{Wang {et~al.}}{2000}]{wcos2000} Wang L., Caldwell R.~R., Ostriker J.~P., Steinhardt P.~J., 2000, ApJ, 530, 17
\bibitem[\protect\citeauthoryear{Will}{2006}]{cw2006} Will C.~M., {\it "The Confrontation between General Relativity and Experiment"}, Living Rev. Relativity 9,  (2006),  3. URL: http://www.livingreviews.org/lrr-2006-3
\bibitem[\protect\citeauthoryear{Zentner}{2007}]{zentner2007} Zentner A.~R., 2007, IJMPD, 16, 763
\bibitem[\protect\citeauthoryear{Zhang \& Hui}{2006}]{zh2006} Zhang J., Hui L., 206, ApJ, 641, 641
\bibitem[\protect\citeauthoryear{Zhao, Li \& Koyama}{2011}]{zlk2011} Zhao G., Li B., Koyama K., 2011, PRD, 83, 044007
\end{thebibliography}
\end{document}